\documentclass[aip,jmp,amsmath,amssymb,superscriptaddress,superbib,twocolumn,groupedaddress]{revtex4}

\usepackage[final]{neel}

\bibpunct{}{}{,}{s}{}{}

\graphicspath{{fig/}}

\usepackage[latin1]{inputenc}
\usepackage[T1]{fontenc}

\def\figurename{Fig.}

\begin{document}

\renewcommand\topfraction{0.8}
\renewcommand\bottomfraction{0.7}
\renewcommand\floatpagefraction{0.7}

\title{Nucleation, imaging and motion of magnetic domain walls in cylindrical nanowires}%

\author{S. Da Col}
\affiliation{Univ. Grenoble Alpes, F-38000 Grenoble, France}%
\affiliation{CNRS, Inst NEEL, F-38000 Grenoble, France}

\author{S. Jamet}
\affiliation{Univ. Grenoble Alpes, F-38000 Grenoble, France}%
\affiliation{CNRS, Inst NEEL, F-38000 Grenoble, France}

\author{M. Sta\v{n}o}
\affiliation{Univ. Grenoble Alpes, F-38000 Grenoble, France}%
\affiliation{CNRS, Inst NEEL, F-38000 Grenoble, France}

\author{B. Trapp}
\affiliation{Univ. Grenoble Alpes, F-38000 Grenoble, France}%
\affiliation{CNRS, Inst NEEL, F-38000 Grenoble, France}

\author{S. Le Denmat}
\affiliation{Univ. Grenoble Alpes, F-38000 Grenoble, France}%
\affiliation{CNRS, Inst NEEL, F-38000 Grenoble, France}

\author{L. Cagnon}
\affiliation{Univ. Grenoble Alpes, F-38000 Grenoble, France}%
\affiliation{CNRS, Inst NEEL, F-38000 Grenoble, France}

\author{J. C. Toussaint}
\affiliation{Univ. Grenoble Alpes, F-38000 Grenoble, France}%
\affiliation{CNRS, Inst NEEL, F-38000 Grenoble, France}

\author{O. Fruchart}
\email[]{olivier.fruchart@cea.fr}
\affiliation{Univ. Grenoble Alpes, F-38000 Grenoble, France}%
\affiliation{CNRS, Inst NEEL, F-38000 Grenoble, France}
\affiliation{CNRS, SPINTEC, F-38000 Grenoble, France}
\affiliation{CEA, INAC-SPINTEC, F-38000 Grenoble, France}

\date{\today}


\begin{abstract}

We report several procedures for the robust nucleation of magnetic domain walls in cylindrical permalloy nanowires. Specific features of the magnetic force microscopy contrast of such soft wires are discussed, with a view to avoid the misinterpretation of the magnetization states. The domain walls could be moved under quasistatic magnetic fields in the range $\unit[0.1\mathrm{-}10]{\milli\tesla}$.

\end{abstract}

\maketitle


\section{Introduction}

The motion of magnetic domain walls~(DWs) in one-dimensional structures has been a subject of increasing interest over the past two decades\cite{bib-THO2007}. Such structures provide a model playground to investigate DW motion under magnetic field or spin-polarized current. They have been proposed to serve as a basis for logic\cite{bib-ALL2002} or memory schemes\cite{bib-PAR2008}. So far, fundamental physics and demonstrators made use of flat strips patterned out of thin films, for ease of fabrication and inspection. One may also consider the cylindrical geometry, which we name wire in the following. Wires can be fabricated with bottom-up techniques by electroplating magnetic metals in insulating templates displaying cylindrical pores\cite{bib-SOU2014}. Dense arrays of vertical wires would be the natural geometry to implement the proposal of a 3D magnetic race-track memory\cite{bib-PAR2008}. For wires, theory and simulations\cite{bib-FOR2002,bib-HER2002a,bib-THI2006} suggested the existence of two types of DWs: the transverse wall and the Bloch-point wall. The features of their motion under field\cite{bib-FOR2002,bib-HER2002a,bib-THI2006} or current\cite{bib-THI2008} were predicted to be even more simple than in strips, mostly precessional in its azimuth in the former case, and purely translational for the second case with speed around $\unit[1]{\kilo\metre\per\second}$, and absence of Walker instabilities.

To search for this physics, there are three requirements: nucleate DWs in a controlled fashion; image them with a simple technique; the material is sufficiently soft so that DWs may be moved under moderate field. These three steps are reported in this manuscript.

\section{Techniques}

As regards synthesis, we start from insulating porous templates obtained by anodization of aluminum in oxalic acid\cite{bib-SOU2014}. Constant voltage leads to straight pores, while for some wires bursts at higher voltage have been used to create local protrusions, \ie with larger diameter. $\mathrm{Fe}_{20}\mathrm{Ni}_{80}$ wires were then obtained by electroplating at $\unit[-1.0]{V}$ versus saturated calomel electrode in an electrolyte containing  $\unit[0.5]{M}$ $\mathrm{Ni}^{2+}$  and $\unit[0.02]{M}$ $\mathrm{Fe}^{2+}$ with $\mathrm{pH}=3$. Finally, the template is dissolved in $\unit[2]{M}$ NaOH , the wires are rinsed several times in water and isopropanol before a drop of solution is left to dry on a supporting surface for further inspection.

Magnetic force microscopy~(MFM) was performed with an NT-MDT NTegra Aura instrument. We used Olympus AC240TS cantilevers (stiffness $\approx\unit[2]{\newton\per\metre}$), custom-coated with $\mathrm{Co}_{80}\mathrm{Cr}_{20}$ of thickness from 3 to $\unit[10]{\nano\metre}$. Imaging was performed in air with the ac two-pass technique, monitoring the phase during the lifted pass. The peak-to-peak amplitude of the tip is circa $\unit[40]{\nano\metre}$. The lift height is in the range $\unit[20\mathrm{-}50]{\nano\meter}$. Micromagnetic simulations were performed with the home-made finite-elements micromagnetic code FeeLLGood\cite{bib-ALO2014}. MFM contrast was estimated as the map of the second vertical derivative of the vertical  component of the simulated stray field.

\section{Nucleation}
\label{sec-nucleation}

Several strategies have been demonstrated to prepare DWs in a controlled manner in patterned strips, such as injection from a large pad\cite{bib-SHI1999b} or nucleation at the bends of curved wires using a large transverse field\cite{bib-TAN1999}. The lesser versatility of design in bottom-up systems makes the preparation of DWs an issue specific to wires.

In a long and narrow wire made of a soft magnetic material, uniform magnetization is the ground state. A non-uniform distribution of magnetization may develop locally at either end such as "C" or curling end domains\cite{bib-FOR2002,bib-HER2002a}. These eventually lead to nucleation of a DW at a value of magnetic field $\Hn$ lower than the average transverse demagnetizing field, the latter being close to $\Ms/2$. Letting aside thermal activation, the nucleation field is determined by the wire diameter normalized to the dipolar exchange length~$\DipolarExchangeLength=\sqrt{2A/(\muZero\Ms^2)}$, with $A$ the exchange stiffness. The value of $\Hn$ decreases for increasing diameter\cite{bib-ZEN2002}. $\Hn$ may equal or exceed $\unit[100]{\milli\tesla}$ for sub-$\unit[50]{\nano\metre}$-diameter wires. Thus, unless the material suffers from very large local pinning~(in which case we would disregard it to investigate DW motion)\cite{bib-HEN2001}, the nucleation of a DW is immediately followed by its very fast propagation along the wire and annihilation at the other end, leaving again the wire in a single-domain state.

One possibility to nucleate and keep a DW at remanence, is to align magnetization exactly perpendicular to the wire axis using a large external field, before going back to remanence. An oscillatory demagnetization along such a direction may be used as well. One or several DWs may be nucleated at a location away from the ends of the wire, thanks to local imperfections or thermal fluctuations. This has been employed successfully by others, and made possible the first observation of a transverse wall in a wire, using electron holography\cite{bib-BIZ2013}. However, our experience is that the alignement of the applied field is critical. Indeed, magnetization is already close to parallel to the wire axis under a transverse field back to $\approx\unit[100]{\milli\tesla}$. Thus, a misalignment of less than one degree is enough to move a created DW towards an end, thus to annihilate it, if the propagation field is of the order of $\approx\unit[1]{\milli\tesla}$. It is desirable to have more robust procedures for nucleation of DWs. We describe below two such procedures that we implemented successfully.

\begin{figure}
  \begin{center}
  \includegraphics[width=72.082mm]{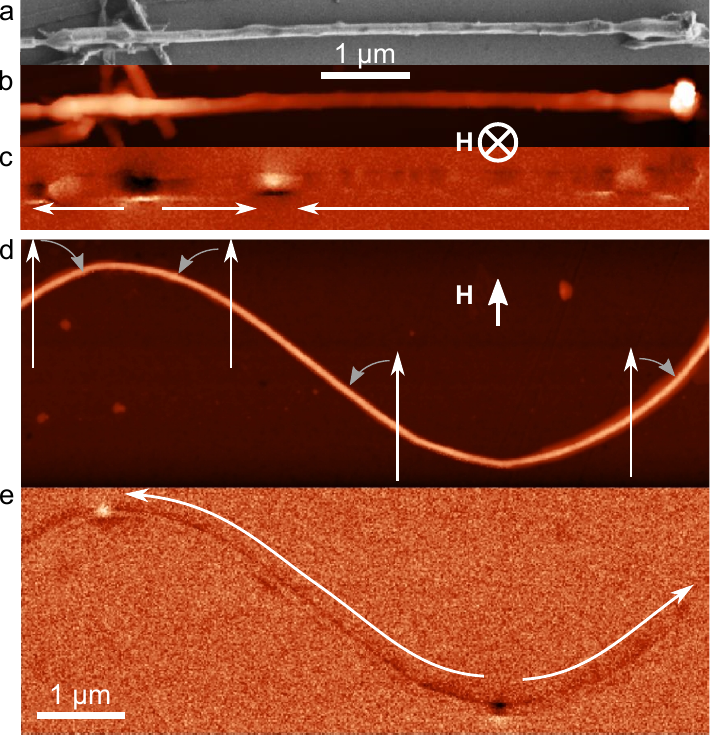}%
  \caption{\label{fig-dw-nucleation}(a)~Scanning electron, (b)~AFM (c)~MFM of a wire with diameter \unit[80]{\nano\meter} displaying local protrusions with diameter~\unit[150]{\nano\meter}, and demagnetized with a large field applied perpendicular to the plane. (d,e)~Atomic and magnetic force microscopy of a bent wire with diameter \unit[60]{\nano\meter}, prepared with an in-plane saturating field.}
  \end{center}
\end{figure}

The first procedure is to modulate the wire diameter along its length. Indeed, as the DW energy increases with the diameter\cite{bib-THI2006,bib-FRU2015b}, protrusions and constrictions are expected to act as energy barriers and wells, respectively. The modulation may induce sufficient pinning so that demagnetization with a transverse magnetic field as described previously, is less critical with the alignment. This procedure is illustrated on \subfigref{fig-dw-nucleation}{(a-c)}, and was used by us to evidence again transverse walls and also for the first time Bloch-point walls by photo-emission electron microscopy\cite{bib-FRU2014}. Note however that, as the resulting pinning field is expected to be lower than the longitudinal nucleation field $\Hn$, a magnetization process with a magnetic field applied along the wire axis still consists of nucleation-propagation-annihilation and cannot lead to a multi-domain wire\cite{bib-PIT2011}.

The second procedure consists in making use of a curved shape, saturating magnetization across the radius, before coming back to remanence. This procedure is straightforward to implement in strips, designed at will by lithography\cite{bib-TAN1999}. As for cylindrical wires, it sometimes happens that a wire is bent during the dispersion procedure. It is possible to drastically increase the yield by first aligning the wires along a given in-plane direction thanks to an applied magnetic field, followed by drying the solution with a blow of air along the in-plane direction transverse to the field. \subfigref{fig-dw-nucleation}{(d,e)} shows a wire prepared this way, where head-to-head and tail-to-tail DWs could be nucleated at two opposite bends.

\section{Imaging}
\label{sec-imaging}

DW motion cannot be tracked with giant magneto-resistance like for strips\cite{bib-GRO2003b}, as a current-in-plane spin-valve structure is not easy to implement for a wire. Magnetic microscopy is therefore expected to play a key role, with MFM being a tool of choice, as an in-lab technique requiring no special sample preparation. However, MFM has known issues, such as tip-sample interaction and image analysis. We discuss below the specific aspects of these in the case of wires.

Tip-sample interaction must be small enough to avoid dragging DWs during imaging, as low-pinning materials are targeted. The stray field of an MFM tip extends over a length scale similar to the radius of curvature of its apex, which is a few tens of nanometers. Thus, for the wires considered here, the interaction issue is \apriori stronger than for flats strips of width several hundreds of nanometers, because the entire DW is under the influence of the stray field. The present MFM images were made using tips with magnetic coating in the range $\thicknm{3\mathrm{-}10}$, which was found to avoid dragging DWs in wires with pinning field as low as $\unit[1]{mT}$. This comes at the expense of a lower MFM signal, with a phase shift much smaller than a degree. On the other hand, such low-moment tips avoid most of the mutual contrast expected to scale like the square of the tip moment\cite{bib-ABR1990}, thus giving a more faithful map of the stray field emanating from the sample.

\begin{figure}
  \begin{center}
  \includegraphics[width=83.31mm]{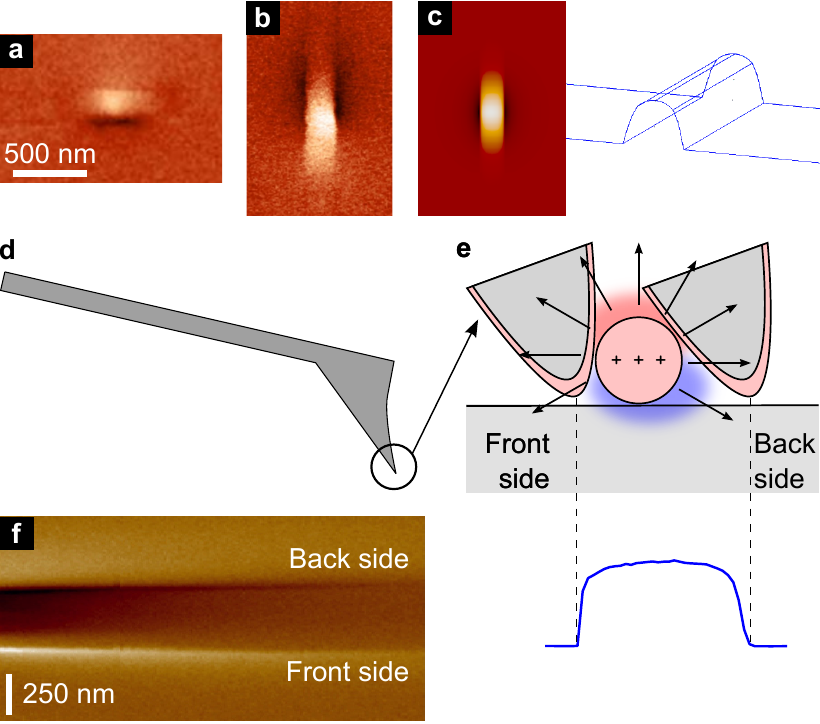}%
  \caption{\label{fig-contrast-analysis}(a,b)~MFM images of a domain wall in a wire with diameter $\unit[70]{\nano\meter}$, with the cantilever perpendicular and parallel to the wire axis, respectively. (c)~Simulated $\linepartialn{H_z}{z}{2}$, based on a Bloch-point wall. The side sketch shows the $\thicknm{10}$-lift surface where the contrast is calculated, reflecting a conical shape for the tip, rounded with a radius of curvature $\thicknm{10}$. The width at the base of the wire was set to $\thicknm{105}$. (d)~Schematic shape of the Olympus tips, and their tilt in our microscope. (e)~Sketch for the stray field associated with magnetic charges in a wire. The bottom part is the cross-section of the experimental topography associated with (f), with a true aspect ratio (f)~Experimental single-line scan close to the end of a $\lengthnm{140}$-diameter wire, while varying the lift height from 20 to $\lengthnm{200}$ from left to right (oscillation amplitude kept constant at \lengthnm{50}).}
  \end{center}
\end{figure}

As regards contrast analysis, in most instruments the tip oscillates essentially along the normal to the sample, and the tip magnetization is also along the same direction. Thus, it is assumed that MFM reflects a vertical derivative of the vertical component of the stray field, itself related to the neighboring sample magnetic charges. Thus, DWs are expected to display a monopolar contrast, dark or light depending on their polarity. To first approximation this is the case in \subfigref{fig-dw-nucleation}{(c,e)}. As already noticed, modulations of diameter\cite{bib-PIT2011} as well as roughness or structural~/ anisotropy fluctuations \cite{bib-IGL2015} also induce a local contrast, with monopolar and dipolar feature along the wire direction, respectively. However, a closer look reveals also a transverse dark/light dipolar contrast perpendicular to the axis at DWs, diameter modulations and wire ends~(\subfigref{fig-dw-nucleation}, and zoom on a DW on \subfigref{fig-contrast-analysis}a). Rotating the wire by $\angledeg{90}$ so that the cantilever is oriented along the wire axis, yields a side contrast again opposite to the contrast on wire, however now symmetric with respect to the wire axis\bracketsubfigref{fig-contrast-analysis}b. The ground for the contrast is the following. When the magnetic center of mass of the tip is below the mid-height plane of the wire, the vertical component of stray field is opposite to that above the wire\bracketsubfigref{fig-contrast-analysis}e. The asymmetric contrast occurring for wires horizontal in the images (\ie transverse to the cantilever direction) stems from the tilt of the tip apex with respect to the sample normal, providing less access to the back side of the wire\bracketsubfigref{fig-contrast-analysis}e. The angle results from the tilt of the cantilever from the sample plane, combined with the tilt of the tip axis with respect to the normal to the cantilever for the Olympus AC series\bracketsubfigref{fig-contrast-analysis}d. Note that the former tilt, which induces a non-vertical direction of oscillation, may also play a role in the front- versus back side contrast, weighing spatial derivatives of the stray field along two directions\cite{bib-RUG1990,bib-FRU2016b}.

\subfigref{fig-contrast-analysis}b is reasonably reproduced by the simulation of MFM contrast of a Bloch-point wall\bracketsubfigref{fig-contrast-analysis}{c}, although not taking into account the above-mentioned tilts. Note also that it is not granted that experimentally a Bloch-point wall may be distinguished from a transverse wall, due to the finite spatial resolution. Anyway, it would be wrong to interpret \subfigref{fig-contrast-analysis}a as a signature for a transverse wall; it is an instrumental feature, which has the strongest signature for large-diameter wires as in \subfigref{fig-dw-nucleation}c compared to \subfigref{fig-dw-nucleation}e, and more generally for small thickness of tip coating, small oscillation amplitude and small lift height\bracketsubfigref{fig-contrast-analysis}f.

\section{Propagation}
\label{sec-propagation}

The wires have been subjected to a quasistatic magnetic field during typically $\unit[1]{s}$. The MFM tip is parked a few micrometers away from the imaging area during the pulse, so that no magnetic bias from the tip is acting on the wire. Imaging is then performed at remanence.

DWs could be moved in the two afore-mentioned cases: wires with either bends or modulations of diameter. In both cases there exists a distribution of pinning sites along the wire length. The distribution of pinning fields in a one-dimensional system due to statistical disorder, has been one of the earliest concepts to describe magnetization reversal. Its general form is now known as the Becker-Kondorski model\cite{bib-BEC1932,bib-KON1937}. It has been detailed recently for wires, evaluating the impact of \eg roughness and fluctuations of magnetocrystalline anisotropy\cite{bib-IVA2011}.

\begin{figure}
  \begin{center}
  \includegraphics[width=72.082mm]{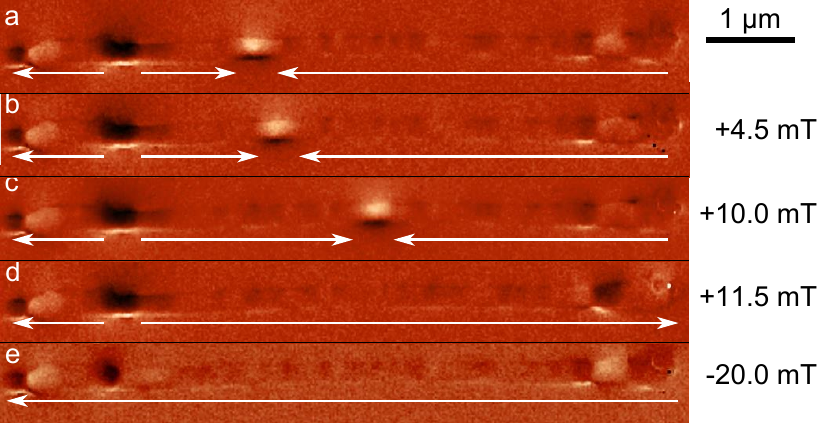}%
  \caption{\label{fig-dw-motion}(a-e)~Successive steps of remagnetization of the wire shown in \subfigref{fig-dw-nucleation}c. The images are taken at remanence, each following a magnetic field with an increasing magnitude from top to bottom. Arrows depict the local direction of magnetization. Positive fields point to the right. In (d) and (e) DWs have moved across modulations of diameter.\dataref{2012-12-03 fils modul\'{e}s AAMS17 Py15.0 dis261112 N}}
  \end{center}
\end{figure}

Examination of several wires reveals a broad distribution of wire- and location-dependent pinning fields, from below $\unit[0.1]{\milli\tesla}$ to around $\unit[10]{\milli\tesla}$. No clear correlation was found with roughness, so that it may result from a material issue involving microstructure and/or strain. Note that such distributions are also a common feature for thin strips deposited by physical means such as sputtering, when investigated step by step by magnetic microscopy\cite{bib-SER2013}.

\section*{Conclusion}

We demonstrated two methods for the controlled nucleation of domain walls (DWs) in cylindrical wires, and highlighted specific features of MFM contrast for such wires. Motion of the DWs is demonstrated with pinning field strength in the range $\unit[0.1\mathrm{-}10]{\milli\tesla}$. These values are similar to those in strips with in-plane or out-of-plane magnetization, in which the dynamics of DW motion was already largely investigated. The route is therefore open to the experimental search of the peculiar features predicted for DW dynamics in cylindrical wires.

\section*{Acknowledgements}

We acknowledge helpful discussions with A.~Wartelle~(Institut NEEL) and A.~Fernandez-Pacheco~(Cavendish Laboratory). This project has received funding from the European Union Seventh Framework Programme (FP7/2007-2013) under grant agreement n$\deg$309589 (M3d).

\section*{References}


%

\end{document}